\documentclass[fleqn,usenatbib]{mnras}
\usepackage{newtxtext,newtxmath}
\usepackage[T1]{fontenc}
\usepackage{ae,aecompl}
\usepackage{graphicx}   
\usepackage{epsfig}
\usepackage{psfig}
\usepackage{color}

\def\gs{\mathrel{\raise0.35ex\hbox{$\scriptstyle >$}\kern-0.6em
\lower0.40ex\hbox{{$\scriptstyle \sim$}}}}
\def\ls{\mathrel{\raise0.35ex\hbox{$\scriptstyle <$}\kern-0.6em
\lower0.40ex\hbox{{$\scriptstyle \sim$}}}}

\makeatother

\date{Accepted ---. Received ---; in original form ---}

\pubyear{2020}

\title[$K$-faint SMGs]{An ALMA survey of the S2CLS UDS field: Optically invisible submillimetre galaxies}

\author[Smail et al.]{Ian Smail,$^{1}$
U.\ Dudzevi\v{c}i\={u}t\.{e},$^{1}$ 
S.\,M.\ Stach,$^{1}$
O.\ Almaini,$^{2}$
J.\,E.\ Birkin,$^{1}$
S.\,C.\ Chapman,$^{3}$
Chian-Chou Chen,$^{4}$\newauthor
J.\,E.\ Geach,$^{5}$
B.\ Gullberg,$^{6}$
J.\,A.\,Hodge,$^{7}$
S.\ Ikarashi,$^{1}$
R.\,J.\ Ivison,$^{8}$
D.\ Scott,$^{9}$
Chris Simpson,$^{10}$\newauthor
A.\,M.\ Swinbank,$^{1}$
A.\,P.\ Thomson,$^{11}$
F.\ Walter,$^{12}$
J.\,L.\ Wardlow$^{13}$ \& 
P.\ van der Werf.$^{7}$
\\
\\
$^{1}$ Centre for Extragalactic Astronomy, Department of Physics, Durham University, South Road, Durham DH1 3LE, UK\\
$^{2}$ School of Physics and Astronomy, University of Nottingham, University Park, Nottingham, NG7 2RD, UK \\
$^{3}$ Department of Physics and Atmospheric Science, Dalhousie University Halifax, NS B3H 3J5, Canada\\
$^{4}$ Academia Sinica Institute of Astronomy and Astrophysics, No.\ 1, Section 4, Roosevelt Rd., Taipei 10617, Taiwan\\
$^{5}$ Centre for Astrophysics Research, Department of Physics, Astronomy \& Mathematics, University of Hertfordshire, Hatfield AL10 9AB, UK\\
$^{6}$ Chalmers University of Technology, Onsala Space Observatory, Onsala, Sweden\\
$^{7}$ Leiden Observatory, Leiden University, P.O.\ box 9513, NL-2300 RA Leiden, The Netherlands\\
$^{8}$ European Southern Observatory, Karl Schwarzschild Strasse 2, Garching, Germany\\
$^{9}$ Department of Physics and Astronomy, University of British Columbia, 6224 Agricultural Road, Vancouver, BC V6T 1Z1, Canada\\
$^{10}$ Gemini Observatory Northern Operations Center, NSF's National Optical--Infrared Astronomy Research Laboratory, 650 North A`oh\={o}k\={u} Place, Hilo Hi 96720, USA\\
$^{11}$ The University of Manchester, Oxford Road, Manchester, M13 9PL, UK\\
$^{12}$ Max-Planck-Institut f\"{u}r Astronomy, K\"{o}nigstuhl 17, 69117 Heidelberg, Germany\\
$^{13}$ Department of Physics, Lancaster University, Lancaster, LA1 4YB, UK\\
}

\begin{document}

\label{firstpage}
\pagerange{\pageref{firstpage}--\pageref{lastpage}}
\maketitle

\begin{abstract}
We analyse a robust sample of 30 near-infrared-faint ($K_{\rm AB}$\,$> $\,25.3, 5$\sigma$)  submillimetre galaxies selected across a 0.96 deg$^2$ field, to investigate their properties and the cause of their lack of detectable optical/near-infrared emission. Our analysis exploits precise identifications based on ALMA 870-$\mu$m continuum imaging, combined with  very deep near-infrared imaging from the UKIDSS-UDS survey.  We estimate that $K_{\rm AB}$\,$>$\,25.3 submillimetre galaxies  represent 15\,$\pm$\,2 per cent of the  total population brighter than $S_{\rm 870}$\,$=$\,3.6\,mJy, with an expected surface density of $\sim$\,450\,deg$^{-2}$ above  $S_{\rm 870}$\,$\geq $\,1\,mJy.  As such they pose a source of contamination in surveys for both high-redshift ``quiescent'' galaxies and very-high-redshift Lyman-break galaxies. We show that these  $K$-faint submillimetre galaxies are simply the tail of the broader submillimetre population, with comparable dust and stellar masses to  $K_{\rm AB}$\,$\leq$\,25.3 mag submillimetre galaxies, but lying at significantly higher redshifts ($z$\,$=$\,3.44\,$\pm$\,0.06 versus $z$\,$=$\,2.36\,$\pm$\,0.11) and having higher dust attenuation ($A_V$\,$=$\,5.2\,$\pm$\,0.3 versus  $A_V$\,$=$\,2.9\,$\pm$\,0.1).  We investigate the origin of the strong dust attenuation and find  indications that these $K$-faint galaxies have smaller dust continuum sizes than the $K_{\rm AB}$\,$\leq$\,25.3   galaxies, as measured by ALMA, which suggests their high attenuation is related to their compact sizes.  We find a  correlation of dust attenuation with star-formation rate surface density ($\Sigma_{\rm SFR}$), with the $K$-faint submillimetre galaxies representing the higher-$\Sigma_{\rm SFR}$ and highest-$A_V$  galaxies. The concentrated, intense star-formation activity in these systems is likely to be associated with the formation of spheroids in compact galaxies at high redshifts, but as a result of their high obscuration these are completely missed in UV, optical and even near-infrared surveys.
\end{abstract}


\section{Introduction}

Dust obscuration is a fundamental characteristic of ultra-luminous infrared galaxies (ULIRGs, with far-infrared luminosities of $L_{\rm IR}$\,$\geq $\,10$^{12}$\,L$_\odot$) at both low and high redshifts, with large columns of dust inferred as a necessary factor to explain the high infrared luminosities and high ratios of far-infrared to optical emission  of these systems (e.g.\ Houck et al.\ 1985; Harwit et al.\ 1987).  High-redshift ULIRGs  are typically uncovered in single-dish sub/millimetre surveys at  mJy-level brightneses, and so are dubbed ``submillimetre'' galaxies (SMG), with lensed examples also frequently found in both far-infrared and longer wavelength studies (e.g.\ Negrello et al.\ 2010; Everett et al.\ 2020).

Observational evidence for optically-faint   (potentially highly dust-obscured or high-redshift) counterparts to SMGs  has been presented since the earliest studies of this population (e.g.\ Rowan-Robinson et al.\ 1991;  Dey et al.\ 1999; Smail et al.\ 1999;  Dunlop et al.\ 2004; Frayer et al.\ 2004; Pope et al.\ 2005; Wang et al.\ 2007, 2011).     Dey et al.\ (1999) suggested that the very red optical-near-infrared colours of  HR\,10, a $K$-band selected submillimetre-detected source at   $z$\,$=$\,1.44, are so extreme   that if a similarly obscured galaxy existed at higher redshifts, $z$\,$\gs$\,3, it would be very challenging to detect in the optical or near-infrared, with $I_{\rm AB}$\,$\gs$\,31 and $K_{\rm AB}$\,$\gs$\,27 (see also Ivison et al.\ 2000; Weiss et al.\ 2009).   Such a optically-invisible, high-redshift SMG was detected in one of the earliest deep submillimetre maps: labelled HDF\,850.1 and subsequently identified using submillimetre inteferometry (Cowie et al.\ 2009; Walter et al.\ 2012), corresponding to  an SMG at $z=$\,5.18, which is undetected in the very deep optical and near-infrared imaging available for this field ($I_{814}$\,$\geq$\,29).   These early results hinted that  high redshift and/or high dust attenuation can easily cause a distant ULIRG to  be undetectable in even the deepest  optical or near-infrared imaging.   Such properties have implications for a range of studies. Firstly, it means that care needs to be taken when using the presence of a near-infrared counterpart as an indication of the reality of a  submillimetre detection (e.g.\ Aravena et al.\ 2016; Dunlop et al.\ 2017), or similarly when searching for potential optical/near-infrared counterparts in larger error circles than justified by the  positional uncertainty. This is true even when using apparently very deep {\it Hubble Space Telescope} ({\it HST}) imaging given their more modest surface brightness sensitivity and  the telescope's inability to operate longward of 1.6\,$\mu$m.  In addition, the extremely red colours of these systems means they represent  a potential source of contamination in studies of high-redshift ``quiescent'' or ``post-starburst'' galaxies (e.g., Simpson et al.\ 2017; Schneider et al.\ 2018) and also for surveys for high-redshift galaxies employing Lyman-break techniques (e.g., Mobasher et al.\ 2005; Pirzkal et al.\ 2013).

The most distant, highly obscured counterparts to submillimetre sources are particularly hard to identify without deep sub/millimetre interferometry (e.g.\ An et al.\ 2019). The  advances over the last decade in the sensitivity of observations in these bands, driven by the Atacama Large Millimeter Array (ALMA) and the upgrades to the  Submillimeter Array (SMA) and the Northern Extended Millimeter Array (NOEMA), have   substantially increased the numbers of these systems available for study.   Simpson et al.\ (2014) analysed early ALMA 870-$\mu$m continuum observations of $\sim$\,100 single-dish submillimetre sources in the ECDFS from  Hodge et al.\ (2013)  and found 19 SMGs that were  undetected in one or more of the {\it Spitzer} IRAC bands between 3.6--8.0\,$\mu$m (half of these SMGs were either undetected in {\it all} four IRAC bands, or had at most a detection in a single band).  They proposed that these near/mid-infrared-faint SMGs lie at somewhat higher redshift, and have either lower stellar masses or are more obscured than the bulk of the optical/near-infrared-detected population (Simpson et al.\ 2014, see also da Cunha et al.\ 2015).    They also highlighted that at least one of these sources is an SMG at $z$\,$=$\,4.4 whose redshift had been identified by Swinbank et al.\ (2012) through the serendipitous detection of the [C{\sc ii}] emission line in their ALMA observations.  A subsequent analysis of the same sample by da Cunha et al.\ (2015),  using the energy-balance SED modelling code {\sc magphys} (da Cunha et al.\ 2008), suggested that the optical/near-infrared faintness of these sources is indeed due to a combination of redshift and dust obscuration.

The completion of interferometric mosaics covering ``blank fields'' (rather than targetting known submillimetre sources detected in panoramic single-dish surveys, as done by Simpson et al.\ 2014) has also turned up a number of optical/near-infrared faint SMGs.  Franco et al.\ (2018) discussed four ALMA-detected sources in the GOODS-S region which lacked $H_{160}$-detections, two of which  are in the earlier  Simpson et al.\ (2014) study (see also Zhou et al.\ 2020).   Yamaguchi et al.\ (2019) found two examples in similar ALMA observations of the same region (one previously identified by Cowie et al.\ 2018), while Umehata et al.\ (2020) obtained a spectroscopic redshift of $z$\,$=$\,3.99 for a near-infrared-faint source uncovered in an ALMA survey in the SSA\,22 region. There have also been examples studied through the selection of sources with very red {\it Herschel} SPIRE colours (e.g.\ Ikarashi et al.\ 2017), and simply through  serendipitous sources discovered in  ALMA observations, e.g.\ Williams et al.\ (2019). Finally, Wang et al.\ (2019), building on previous work that connected galaxies having extremely red near-infrared colours with submillimetre sources (e.g.\ Smail et al.\ 1999; Im et al.\ 2002; Frayer et al.\ 2004; Coppin et al.\ 2004), obtained ALMA snapshot continuum observations of a sample of 63 sources with very red $H_{160}-m_{4.5}$ colours and detected 39 of these with 870-$\mu$m flux densities, $S_{\rm 870}$, of $S_{\rm 870}$\,$\geq$\,0.6\,mJy.   

In this work we exploit a large ALMA SMG survey by Stach et al.\ (2019), who catalogued 708 SMGs from
  870-$\mu$m interferometric continuum follow-up observations of 716 SCUBA-2 submillimetre sources in  the SCUBA-2 Cosmology Legacy Survey (S2CLS, Geach et al.\ 2017) map of the UKIDSS Ultra Deep Survey (UDS, Lawrence et al.\ 2007; Almaini et al.\ in prep.).    Dudzevi\v{c}i\={u}t\.{e} et al.\ (2020) published an analysis of the multiwavelength properties of these ALMA-identified SMGs.   Here we focus on the subset of these SMGs that are faint or undetected in the very deep $K$-band imaging ($K_{\rm AB}$\,$=$\,25.3, 5\,$\sigma$) obtained for the UDS region by Almaini et al.\ (in prep.), and use this statistically robust sample to investigate the nature of these very faint and very red SMGs.  We assume a cosmology with $\Omega_{\rm M}$\,$=$\,0.3, $\Omega_\Lambda$\,$=$\,0.7 and $H_0$\,$=$\,70\,km\,s$^{-1}$\,Mpc$^{-1}$, all quoted magnitudes are on the AB system and errors on median values are derived from bootstrap resampling.  In this cosmology at the median redshift of our $K_{\rm AB}$\,$>$\,25.3 SMG sample, $z$\,$=$\,3.4, 1 arcsec corresponds to 7.5\,kpc.

%
%
\setcounter{figure}{0}
\begin{figure*}
\centerline{\psfig{file=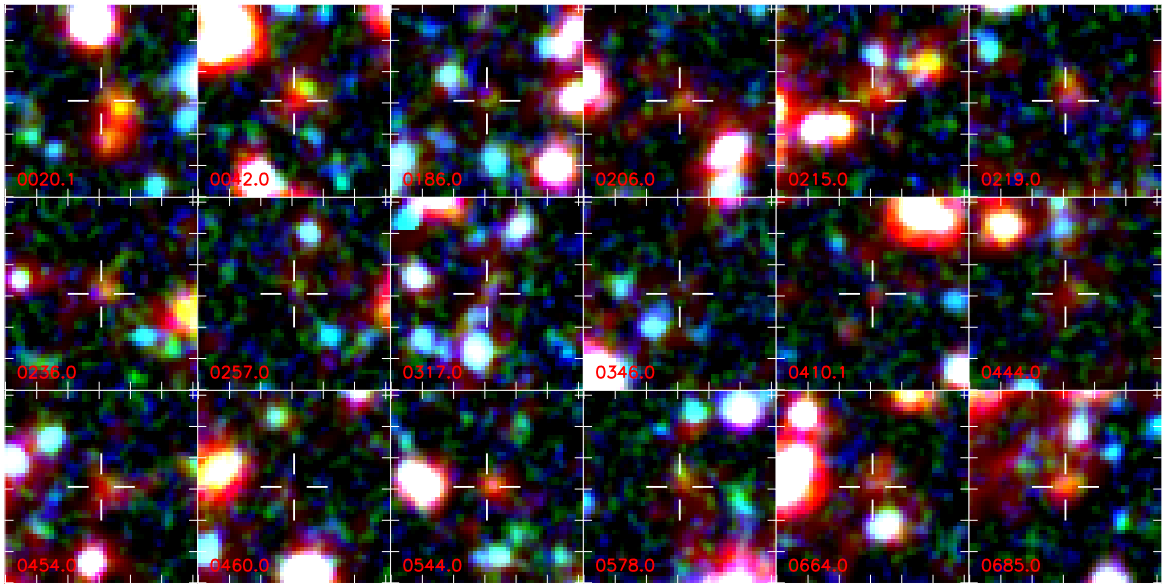,width=3.in,angle=90}}
\caption{\small 
False colour images ($J$, $K$ and 3.6+4.5\,$\mu$m as the blue, green and red channels respectively) centred on example SMGs from our {\it K-faint} sample, with the cross-hair marking the position of the ALMA source in each field. The typically very red near-infrared colours of these galaxies, as well as their faintness, are clear. Each panel is  12 arcsec square and the $J$- and $K$-band images have been smoothed with 0.5-arcsec FWHM Gaussians to enhance the visibility of faint emission.
}
\end{figure*}

\section{Observations and Analysis}

Our analysis is based on the multi-band photometric catalogue of 707 ALMA-located SMGs in the UKIDSS UDS field, constructed by Dudzevi\v{c}i\={u}t\.{e} et al.\ (2020) from the original ALMA catalogue in Stach et al.\ (2019). We  provide a brief summary of the catalogue here and we refer the reader to those papers for the full details.

\subsection{Sample selection}

Stach et al.\ (2019) obtained ALMA Band 7 continuum observations in Cycles 1, 3, 4 and 5 of a complete sample of 716  $>$\,4-$\sigma$ ($S_{\rm 850}$\,$\geq$\,3.6\,mJy) single-dish submillimetre sources selected from the 0.96-deg$^2$ S2CLS 850-$\mu$m map of the UDS field (Geach et al.\ 2017).   Stach et al.\ (2019) catalogued 708 sources within the primary beam areas of these observations above a threshold of 4.3\,$\sigma$ (corresponding to a false-positive rate of two per cent) based on maps tapered to a uniform resolution of 0.5-arcsec FWHM.  The 707 SMGs have fluxes of $S_{\rm 870}$\,$=$\,0.6--13.6\,mJy, after removing the brightest source, which corresponds to a strongly-lensed SMG  identifed by Ikarashi et al.\ (2011).    

Dudzevi\v{c}i\={u}t\.{e} et al.\ (2020) utilised the multi-band imaging of the UDS field gathered by Almaini et al.\ (in prep.) to construct broadband spectral energy distributions (SEDs) of the ALMA sources.   This imaging includes the DR11 UKIRT WFCAM observations of the UDS in $JHK$, reaching  5-$\sigma$ depths of $J$\,$=$\,25.6, $H$\,$=$\,25.1 and $K$\,$=$\,25.3, as well as deep $UBVRi'z'Y$ imaging from Subaru, VISTA and CFHT.  In addition, deep {\it Spitzer} IRAC/MIPS 3.6--24\,$\mu$m coverage is provided by the SpUDS survey (PI: J.\ Dunlop),  1.4-GHz radio observations come from UDS20 survey (Arumugam et al.\ in preparation; for a summary see Simpson et al.\ 2013), while  {\it Herschel} SPIRE 250--500\,$\mu$m photometry was deblended using the ALMA sources (as well as radio and MIPS priors) following Swinbank et al.\ (2014).  Dudzevi\v{c}i\={u}t\.{e} et al.\ (2020)  employed the {\sc magphys} photo-$z$ energy balance  modelling code (da Cunha et al.\ 2015; Battisti et al.\ 2019) to fit this multiband photometry.   Derived parameters include photometric redshifts ($z$), far-infared luminosities (8--1000\,$\mu$m, $L_{\rm IR}$), dust masses ($M_{\rm d}$) and $V$-band attenuation ($A_V$), see Dudzevi\v{c}i\={u}t\.{e} et al.\ (2020) for more details of the models and the extensive tests which were applied using both observed and simulated data.   We note that as we are studying a population whose SEDs include significant numbers of filters with no detections we have tested the sensitivity of the photometric redshifts to changing the definitions of these limits. We vary the expected fluxes to be either 0 or 1.5\,$\sigma$ in either the wavebands shortward of an observed wavelength of 8\,$\mu$m (in the optical/near-infrared), or longward (in the far-infrared/submillimetre), or both.  In this way we confirm that the offsets are no larger than  $\sim $\,0.05 in $\Delta z/(1+z)$ with a scatter of $\sim$\,0.1--0.2 (see also Dudzevi\v{c}i\={u}t\.{e} et al.\ 2020).     

To isolate a sample of $K$\,$>$\,25.3 SMGs we first apply a signal-to-noise threshold (in a 0.5-arcsec aperture) of $\geq$\,4.8\,$\sigma$. This threshold corresponds to a false positive rate of 1 source in the sample 637 SMGs with SNR\,$\geq$\,4.8 across the $\sim$\,50\,arcmin$^2$ combined area mapped with ALMA, based on the detection rate of false sources in the inverted maps (Stach et al.\ 2019).  We choose this more conservative threshold since we are studying the properties of   the near-infrared faint subset of the SMG population, that would otherwise potentially suffer significant contamination from spurious ALMA sources if we allowed a higher false positive rate.   We then restrict the sample to those sources falling within the footprint of the  very deep UKIDSS DR11 UDS $K$-band observations and having coverage in more than 10 broadband filters in the optical/near-infrared, including observations in all four {\it Spitzer} IRAC channels.  This provides an initial sample of 496 SMGs (271 of these have $S_{870}$\,$\geq\,$\,3.6\,mJy, the flux density limit of the parent SCUBA-2 survey of the UDS), of which 80 do not have a catalogued $K$-band counterpart brighter than the 5-$\sigma$ limit of the UDS $K$-band imaging  $K$\,$=$\,25.3 mag in a 2-arcsec diameter aperture.  We adopt this $K$-band limit as it is broadly representative of the detection limit for extended sources in current ground- and space-based near-infrared surveys  (e.g.\ $H_{160}$\,$\sim$\,26 in the CANDELS WFC3 imaging used by Franco et al.\ 2018).
 
We then visually checked these sources in the DR11 $K$-band image (rebinned 2\,$\times$\,2 to 0.26 arcsec pixel$^{-1}$ sampling to increase the visibility of faint sources) to remove those with bright nearby galaxies within a 5-arcsec diameter region, which may contaminate the optical/near-infrared photometry (some of these may also be  gravitationally lensing the SMG, although only by modest factors), or ambiguous cases where a faint $K$-band counterpart lies within 2 arcsec, where the offset between the submillimetre and near-infrared emission could plausibly be due to different levels of dust obscuration within a single galaxy.   This gives a sample of 30 {\it K-faint} SMGs where we are confident that the 2-arcsec diameter photometry of the submillimetre sources is not contaminated by other nearby galaxies.   We focus our analysis on this {\it K-faint} subset, while noting that they are expected to be broadly representative of the larger full sample of 80 $K$\,$>$\,25.3 SMGs, but with more robust photometry.   These 30 SMGs are AS2UDS IDs 0020.1, 0024.0, 0042.0, 0058.0, 0137.1, 0161.0, 0186.0, 0205.0, 0206.0, 0215.0, 0219.0, 0236.0, 0257.0, 0309.0, 0310.0, 0317.0, 0346.0, 0355.0, 0369.0, 0410.1, 0444.0, 0454.0, 0460.0, 0544.0, 0564.0, 0578.0, 0639.0, 0664.0, 0685.0 and 0690.0.   Five of these fall in the CANDELS WFC3 coverage of the UDS and we have confirmed that none of these are detected above $H_{160}$\,$\sim$\,26.0--26.3 (5\,$\sigma$).

In addition, we construct a control sample from the $K$-detected SMGs, selected to have
the same photometric redshift distribution and $L_{\rm IR}$ as the $K$-blank sample, to allow comparisons free of the evolutionary (and selection) trends seen in the broader population (e.g.\ Stach et al.\ 2019; Dudzevi\v{c}i\={u}t\.{e} et al.\ 2020).  We do this by searching around each of the SMGs in the {\it K-faint} sample  for the  nearest  SMGs in the $K$-detected sample within $\Delta$\,$=$\,0.1 in both $z$ and $\log_{10} L_{\rm IR}$.   Removing duplicate matches, this leaves us with a sample of 100 $K$-detected SMGs in our {\it ``control''} sample, which are matched in redshift and $L_{\rm IR}$ to the 30 {\it K-faint}  SMGs.

%
%
\begin{figure*}
\centerline{
\psfig{file=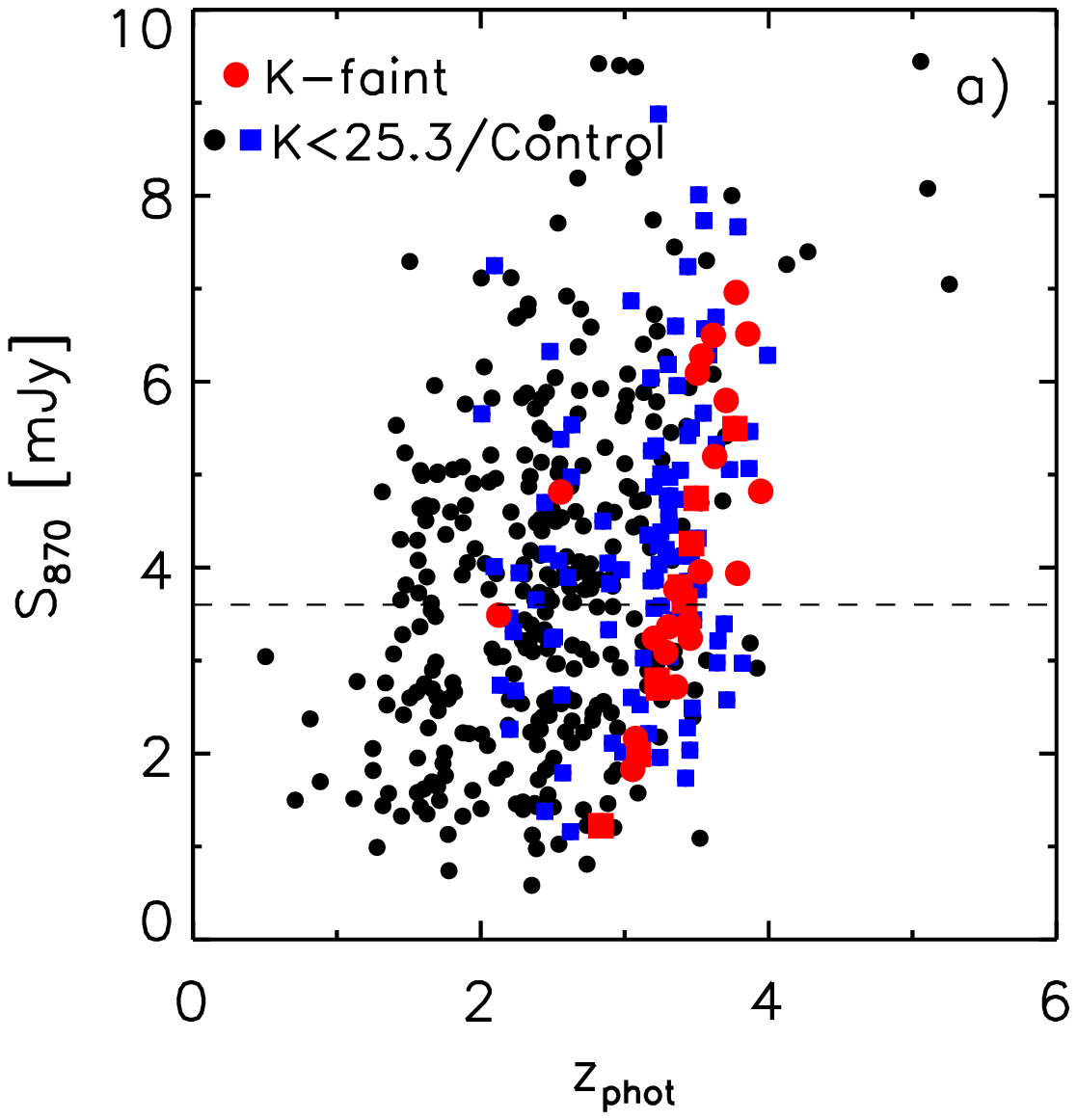,width=2.5in,angle=0} 
\hspace*{-0.5in}
\psfig{file=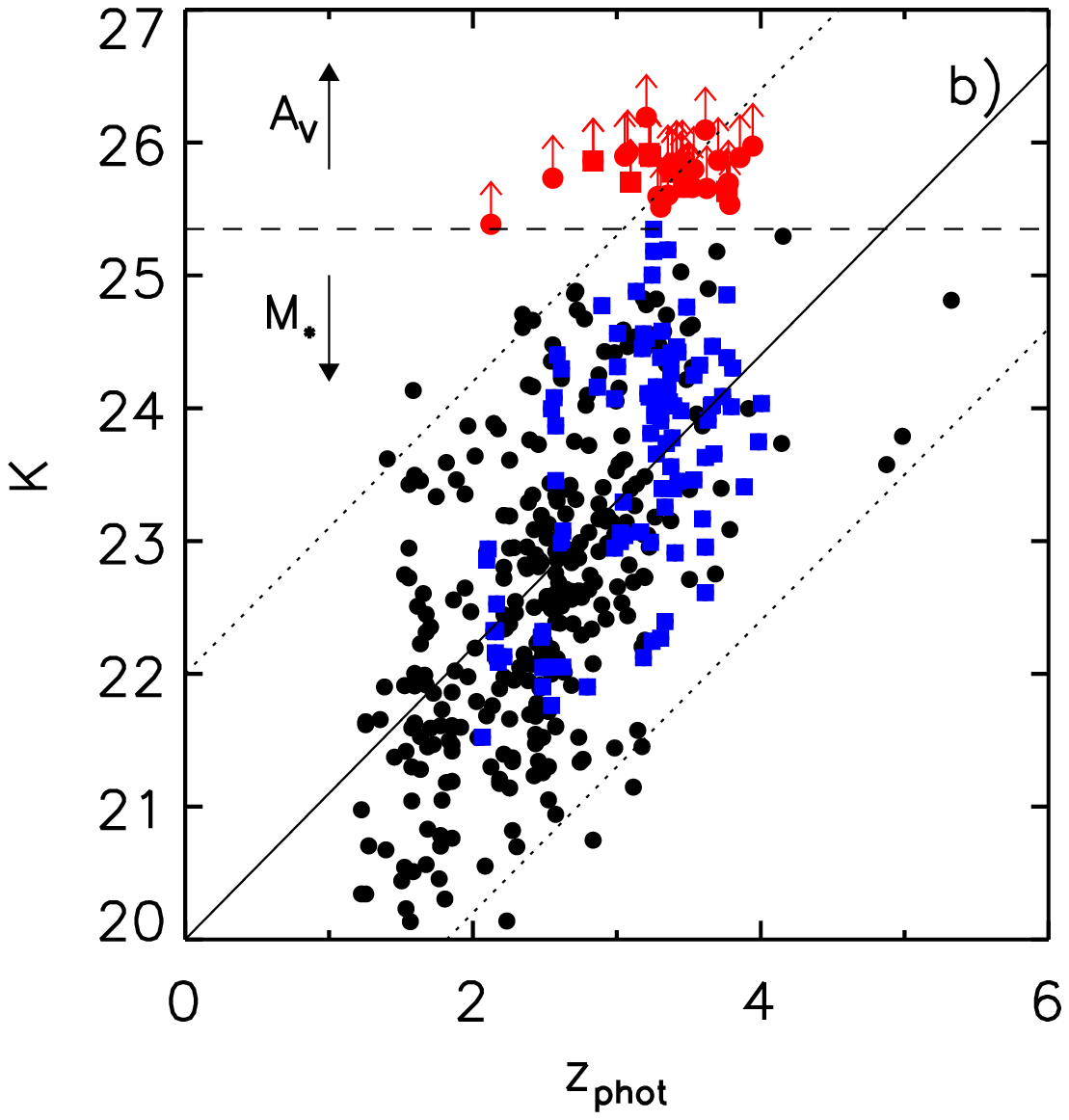,width=2.5in,angle=0} 
\hspace*{-0.3in}
\psfig{file=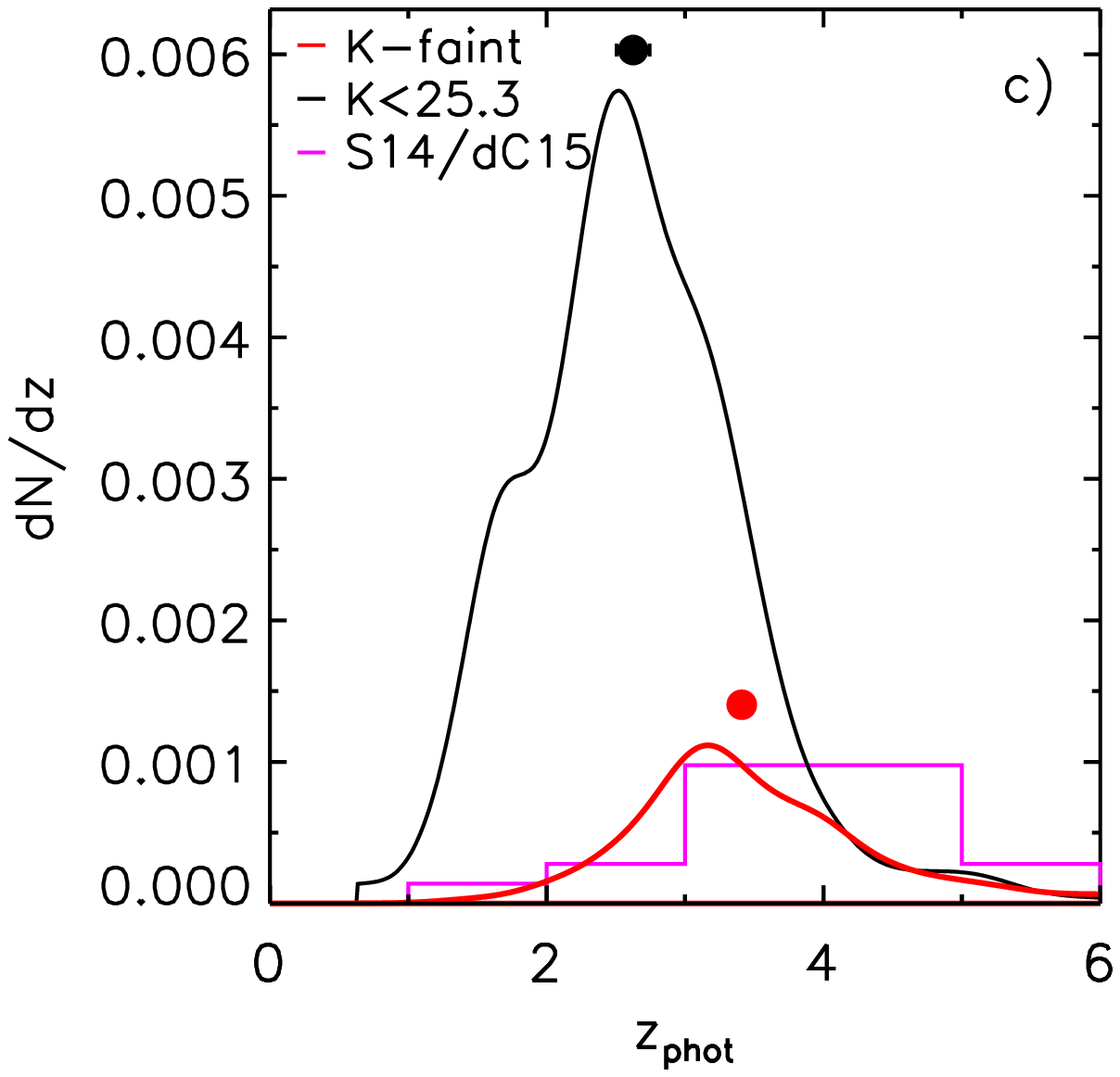,width=2.5in,angle=0}
}
\caption{\small 
{\it a)} Submillimetre flux density ($S_{870}$) versus median photometric redshift for the {\it K-faint} $K$\,$>$\,25.3 sample, compared to the $K$\,$\leq$\,25.3 SMGs in AS2UDS, where we highlight the sources in the $z/L_{\rm IR}$-matched {\it control} sample.   We see that the {\it K-faint} sources typically lie at higher redshifts than the majority of the $K$-detected SMGs, but that even at $z$\,$\gs$\,3--4 there is a mix, with the {\it K-faint} sample comprising at most $\sim$\,25--30 per cent of the SMG population.    We  identify the nine {\it K-faint} SMGs whose SEDs are  constrained solely by photometric limits by plotting those as squares.  
{\it b)} $K$-band magnitude versus photometric redshift for the $K$\,$\leq$\,25.3 SMGs in AS2UDS and the  {\it K-faint} sample (shown as a scatter of points fainter than the 5-$\sigma$ detection limit of $K$\,$=$\,25.3, shown by the dashed line).  The vectors indicate the influence of increasing stellar mass ($M_\star$) or attenuation ($A_V$).
We show a linear fit to the median trend line as a solid line and the dotted lines offset by $\pm$\,2\,magnitudes which roughly delimit the boundaries of the distribution. Given the trends seen in the population we conclude that the $K$-band magnitudes of the {\it K-faint} SMGs are consistent with being both higher redshift, but also either less massive or more dust attenuated.
{\it c)} The redshift distribution and median redshifts for the {\it K-faint} SMG sample compared to the $K$-detected AS2UDS SMGs and the distribution of near-infrared-faint SMGs from Simpson et al.\ (2014).   The distributions for the AS2UDS samples are the average {\sc magphys} PDFs from Dudzevi\v{c}i\={u}t\.{e} et al.\ (2020), while we show the binned {\sc magphys}  photometric redshifts from  da Cunha et al.\ (2015) for the Simpson et al.\ (2014) sources (S14/dC15).  We confirm the behaviour seen in  panel a), that the {\it K-faint} SMGs have a significantly higher median redshift than the $K$-detected sample: $z$\,$=$\,3.44\,$\pm$\,0.06 (consistent with $z=$\,3.8\,$\pm$\,0.4 for the Simpson et al.\ 2014 examples) and $z$\,$=$\,2.36\,$\pm$\,0.11, respectively (close to the values reported for similar selections by Dudzevi\v{c}i\={u}t\.{e} et al.\ 2020). These estimates are based on the median of the individual redshifts for each distribution and we plot these and indicate their bootstrap uncertainty on the figure. The {\it K-faint} distributions have been normalised to reflect the number densities of their parent samples.}
\end{figure*}

\subsection{Photometric properties}

The median submillimetre flux density for the {\it K-faint} sample of 30 SMGs is $S_{\rm 870}$\,$=$\,3.8\,$\pm$\,0.3\,mJy (which is identical to the median of the whole sample of 496 SMGs).  Eleven of these 30 SMGs are detected in the {\it Herschel} SPIRE or PACS bands, or at 1.4\,GHz with the VLA (see Dudzevi\v{c}i\={u}t\.{e} et al.\ 2020), reflecting their typically modest submillimetre brightness and the expected higher-than-average redshifts of this sample.  Of the three radio detected {\it K-faint} SMGs, the most noteworthy is AS2UDS\,0454.0, which is a radio-loud AGN with $S_{\rm 1.4}$\,$=$\,0.6\,mJy at $z$\,$=$\,3.5$_{-0.7}^{+0.8}$ (see Algera et al.\ 2020).

Figure~1 shows false-colour images of SMGs from  our sample using the  $J$, $K$ and 3.6+4.5\,$\mu$m passbands.  We see that, as expected, all of the galaxies are both very faint in the near/mid-infrared and when detected they display very red colours.   Where detected in the IRAC bands, the galaxies are typically compact in this $\sim$\,2-arcsec FWHM imaging.  Since we are studying {\it K-faint} SMGs,  these are neither detected  individually in the higher resolution $K$-band imaging, nor in a $K$-band stack (see below), with sufficient signal to noise to measure a reliable size, although the stacked source sizes are not inconsistent with the $R_{\rm e}$\,$=$\,4.4$^{+1.1}_{-0.5}$\,kpc reported in the $H_{160}$-band by Chen et al.\ (2015) for $H$-band detected SMGs at $z$\,$=$\,1--3.   

To place limits on the  characteristic near-infrared brightness of these systems we stack the sample of 30 SMGs in the UDS $J$- and $K$-band imaging  (Almaini et al.\ in prep.) and obtain a marginal detection in the $K$-band, with an average magnitude of $K$\,$=$\,26.0\,$\pm$\,0.5, with the sources undetected in the $J$ band, corresponding to a 3-$\sigma$ limit of $J$\,$\gs$\,27.5.  Of the 30 SMGs in the {\it K-faint} sample, 16 are detected in one or more of the {\it Spitzer} IRAC channels at 3.6--8.0\,$\mu$m, with the median magnitudes of these detected sources being  $m_{\rm 3.6}$\,$=$\,23.3\,$\pm$\,0.2,  $m_{\rm 4.5}$\,$=$\,22.9\,$\pm$\,0.3,  $m_{\rm 5.8}$\,$=$\,22.1\,$\pm$\,0.1 and $m_{\rm 8.0}$\,$=$\,22.1\,$\pm$\,0.1.  We stack the IRAC imaging of the remaining 14 undetected SMGs  and obtain a weak detection at 4.5\,$\mu$m and limits in the other bands:  $m_{\rm 4.5}$\,$=$\,24.7\,$\pm$\,0.3,    $m_{\rm 3.6}$\,$\geq$\,24.9,   $m_{\rm 5.8}$\,$\geq$\,23.5 and $m_{\rm 8.0}$\,$\geq$\,23.5 (3-$\sigma$ limits).  The IRAC 4.5-$\mu$m band is the reddest deep band and so it is unsurprising that we detect the sources in this filter and not in the bluer 3.6-$\mu$m channel or the shallower 5.8- or 8.0-$\mu$m channels.    The 14 {\it K-faint} SMGs that lack individual IRAC detections are the most challenging sources to study, owing to the absence of constraints on their broadband SEDs.  Nevertheless, we note that of the 14, five are detected in SPIRE, PACS or radio bands, leaving only nine with no additional constraints on their SEDs other than limits.   We stress that our adoption of a $\geq$\,4.8\,$\sigma$ ALMA selection means that at most  one of these  sources is expected to be spurious.   The median ALMA flux density of these nine SMGs is  consistent with the full sample,  $S_{\rm 870}$\,$=$\,3.6\,$\pm$\,0.6\,mJy, and in addition two of them are independently detected by Ikarashi et al.\ (2015).   One of these, AS2UDS\,0186.0 is one of the two SMGs with very faint near-infrared counterparts and compact dust continuum emission studied by Ikarashi et al.\ (2017),  ASXDF1100.053.1, who suggested the very red  far-infrared/submillimetre colours of these galaxies indicated they lay at high redshifts, $z$\,$\gs$\,4.

\subsection{Structural properties}

In addition to the parameters derived from the SED modelling, our analysis also exploits dust continuum sizes for a subset of the AS2UDS sample from Gullberg et al.\ (2019).  This size information is available because a large fraction of the multi-cycle ALMA observations taken for the survey were obtained with the array in moderately extended configurations, yielding synthesised beams with angular sizes of $\sim$\,0.2\,arcsec, sufficient to reliably resolve the dust emission in those sources detected at sufficient signal to noise (SNR, SNR\,$\geq$\,8).  Gullberg et al.\ (2019) present circularised effective radii, $R_{\rm e}$, for 153 SMGs detected at SNR\,$\geq$\,8, from fitting Sersic $n=$\,1 profiles to the 870-$\mu$m continuum images.\footnote{We caution that the values reported for the circularised $n=$\,1 effective radii in Table~A1 in Gullberg et al.\ (2019), $R^{\rm A1}_{\rm e}$,  are in error and actually list the minor-axis size, the true $n=$\,1 circularised $R_{\rm e}$ can be easily recovered using the Axial Ratio ($b/a$) values in the table: $R_{\rm e}$\,$=$\,$\sqrt{(R^{\rm A1}_{\rm e})^2/(b/a)}$.}    

We highlight two critical facts about these observations:  firstly, while the sources are well-fit by exponential surface brightness profiles, the ellipticity distribution of the sources suggests that the dust continuum emission arises from tri-axial structures, most likely bars (see Gullberg et al.\ 2019).  This supports the conclusions from deeper, high-resolution imaging of a smaller sample of SMGs by Hodge et al.\ (2016, 2019) which directly resolves bar-like structures in those sources.   The second issue to note is that the shallow AS2UDS observations only detect the highest surface brightness components,  with Gullberg et al.\ (2019) showing from a stacking analysis that in addition to these high-surface brightness, compact components (with $R_{\rm e}$\,$\sim$\,0.1\,arcsec or $\sim$\,1\,kpc), a typical source also has a fainter, more extended exponential component ($R_{\rm e}$\,$\sim$\,0.5\,arcsec of $\sim$\,4\,kpc) which is undetected in the individual maps.   This extended component has a size comparable to the stellar and gas disks in these systems (Gullberg et al.\ 2019; see also Calistro Rivera et al.\ 2018). To account for the contribution of this faint, extended component to the sizes we statistically correct  the sizes  using the flux-weighted sum of the measured size in Gullberg et al.\ (2019) and a 0.5-mJy component with a size of 0.5$''$, as indicated by their stacking analysis.  This accounts for the poor sensitivity of the high-resolution ALMA snapshot observations to the faint, extended emission in these sources. The {\it corrected} radius is then $R^{\rm corr}_e$\,$=$\,$(S_{870}-0.5)/S_{870} \times R^{\rm true}_{\rm e} + 0.5/S_{870} \times 0.5''$.  The correction is a median of factor of 1.30$^{+0.25}_{-0.13}$, where the error is the 16--84th percentile range, but we stress that this correction, while reproducing the average true sizes, will not recover the full dispersion of the true sizes.

In our analysis we also include 870-$\mu$m sizes for nine $K$-bright SMGs which were also observed by Tadaki et al.\ (2020) in their deeper, high-resolution ALMA survey of a $K$-band selected galaxies including the UDS field.  A further two SMGs  from Gullberg et al.\ (2019) have sizes reported by Tadaki et al.\ (2020) and these agree with the corrected values used here.  We also employ the 42 remaining fainter UDS galaxies  from Tadaki et al.\ (2020) with submillimetre detections at SNR\,$\geq$\,10 and measured sizes as a comparison sample in our work. To do this we use {\sc magphys} to model their multiwavelength photometry, including the reported $S_{870}$ fluxes, in an identical manner to that used for the AS2UDS sample by Dudzevi\v{c}i\={u}t\.{e} et al.\ (2020).

%
%
\begin{figure*}
\centerline{
\psfig{file=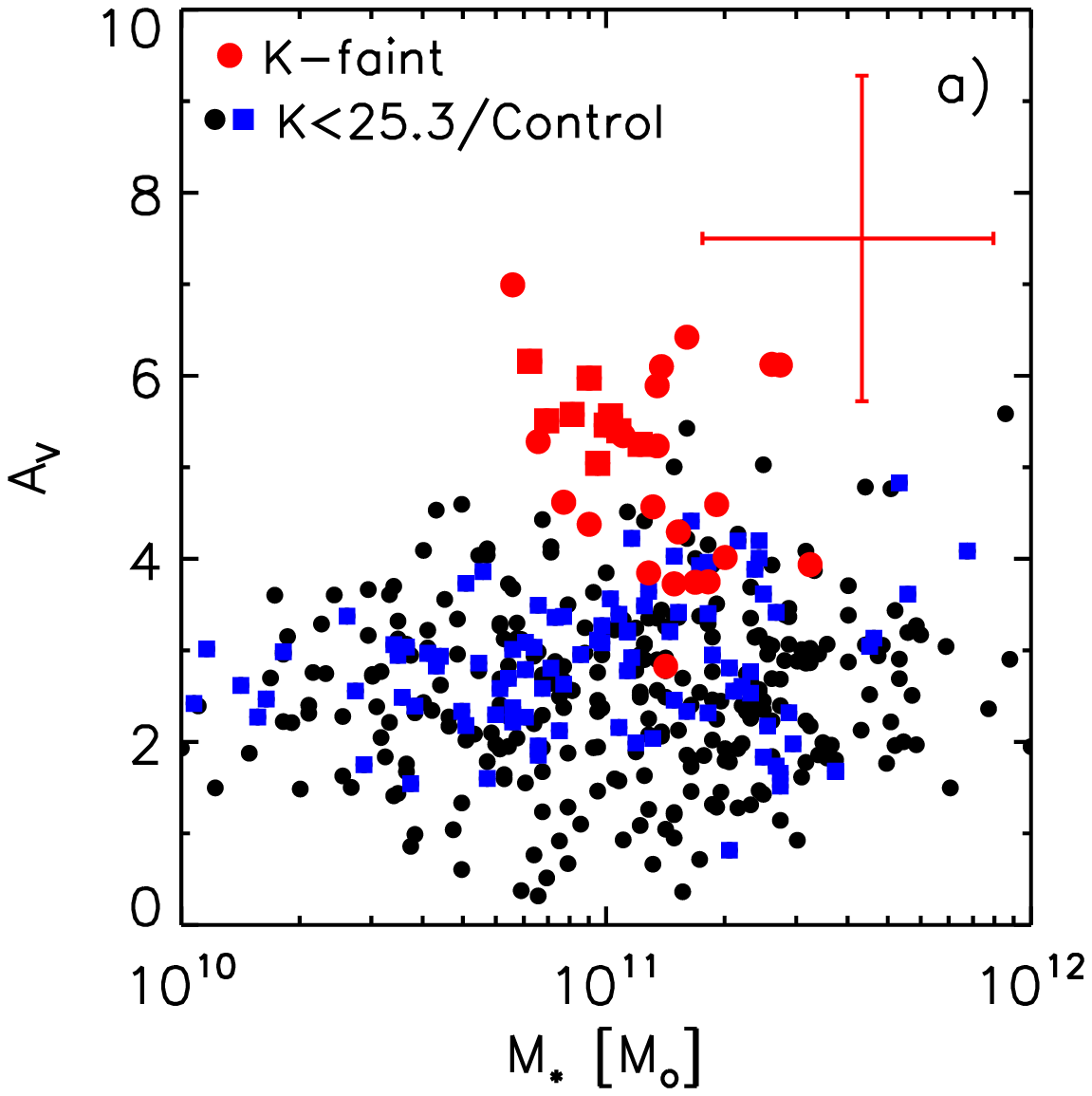,width=3.in,angle=0}
\hspace*{-0.8in}
\psfig{file=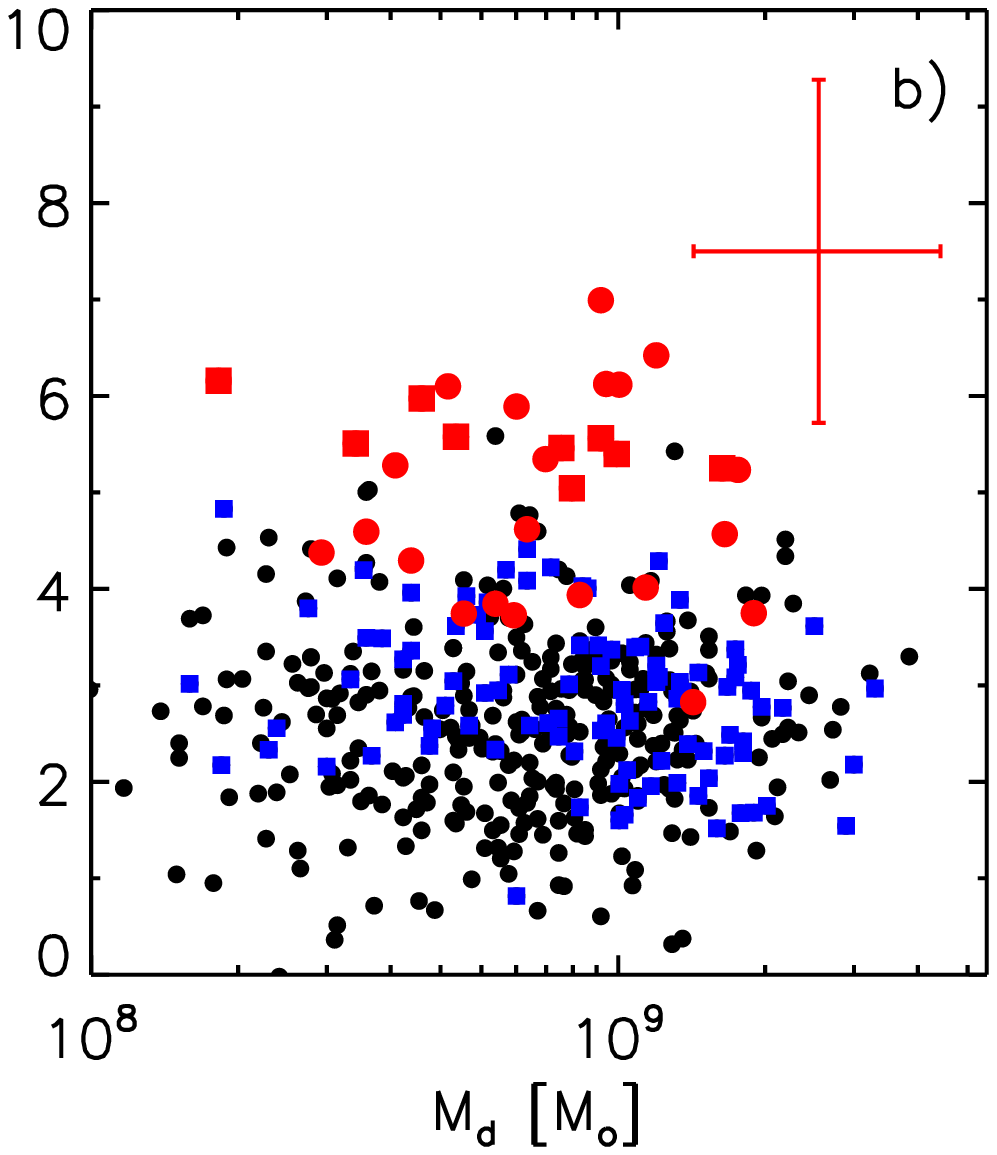,width=3.in,angle=0}
\hspace*{-0.8in}
\psfig{file=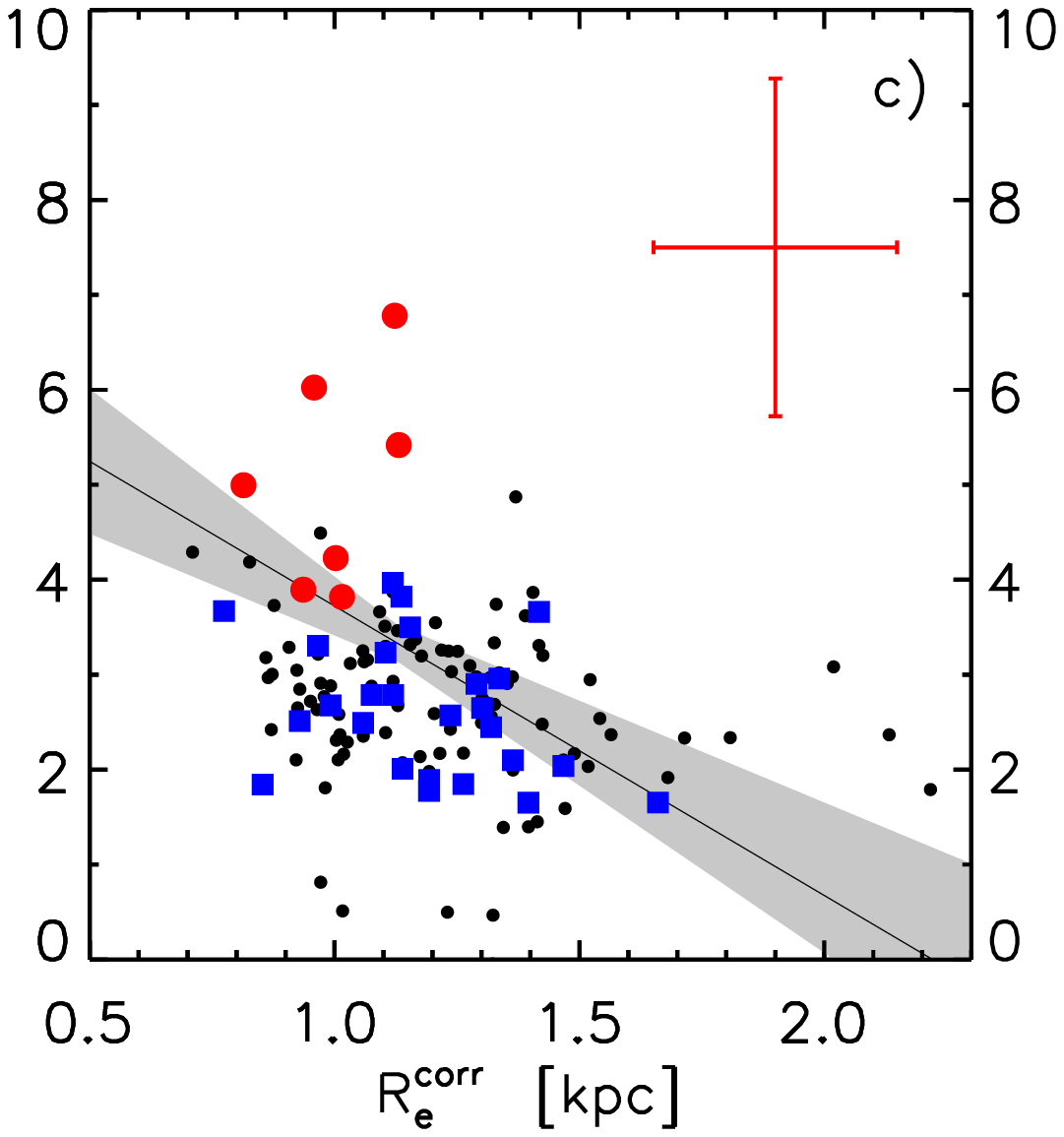,width=3.in,angle=0}}
\caption{\small 
{\it a)}   Variation of $V$-band attenuation ($A_V$) with stellar mass ($M_\ast$) for the {\it K-faint}SMGs, compared to the population of $K$-detected SMGs with $K$\,$\leq$\,25.3 mag.    We see that the {\it K-faint} SMGs have higher $A_V$, as expected, but span only a modest range in stellar mass compared to the broader population (or the {\it control}), with typically high stellar masses of $M_\ast$\,$=$\,10$^{11.10\pm 0.04}$\,M$_\odot$. We highlight the sources in the $z/L_{\rm IR}$-matched {\it control} sample of the $K$-detected SMG population and we similarly identify the nine {\it K-faint} SMGs whose SEDs are  constrained solely by photometric limits by plotting those as squares.   
{\it b)}   Variation of $A_V$ with dust mass ($M_{\rm d}$) for the {\it K-faint} SMGs compared to the sample of $K$-detected SMGs.   Here we see that the {\it K-faint} SMGs exhibit a broad range in $M_{\rm d}$, comparable to that of the $K$-detected and {\it control} samples.
{\it c)} Variation of $A_V$  with dust continuum size ($R_{\rm e}$) for the subset of SMGs that have measurements from Gullberg et al.\ (2019), which shows a  correlation for the combined {\it control} and {\it K-faint} samples, with the {\it K-faint} SMGs typically having smaller-than-average $R_{\rm e}$, as well as higher-than-average $A_V$. We overplot a linear fit to the  trend seen in the combined sample as a solid line and illustrate the 1-$\sigma$ uncertainties in this by the  shaded region.   In each panel we illustrate the median fractional uncertainty in the parameters for the sources in the {\it K-faint} sample, based on the 16--84\,th percentile range of the relevant PDF from the {\sc magphys} analysis (see Dudzevi\v{c}i\={u}t\.{e} et al.\ 2020).
}
\end{figure*}

\section{Results and Discussion}

We start by  assessing the rate of $K$-faint SMGs in the overall SMG population.  We adopt an ALMA flux density limit of  $S_{\rm 870}$\,$\geq$\,3.6\,mJy to ensure that the sample is complete over the UDS field (see Stach et al.\ 2019), finding a lower limit of 17 SMGs from a parent population of 271 SMGs using our {\it K-faint}  sample ($\geq$\,6\,$\pm$\,2 per cent) and 42 examples (15\,$\pm$\,2 per cent) from the full $K$\,$>$\,25.3 sample, which are consistent with the 17\,$\pm$\,1 per cent reported by Dudzevi\v{c}i\={u}t\.{e} et al.\ (2020).   We therefore conclude that 15\,$\pm$\,2 per cent of SMGs brighter than $S_{\rm 870}$\,$\geq$\,3.6\,mJy are near-infrared faint ($K$\,$\gs$\,25.3 mag).   We see no significant variation in the fraction of $K$-faint SMGs with 870-$\mu$m flux density across the range $S_{\rm 870}$\,$=$\,3--10\,mJy. 

Our  measured $K$-faint SMG fraction is consistent with  previous estimates of the fraction of $K$-faint sources in the SMG population based on smaller samples:  20\,$\pm$\,4 per cent from the 19 sources in the Simpson et al.\ (2014) sample fainter than $K$\,$\sim$\,24.4 (3$\sigma$);  $\sim$\,20 per cent by Franco et al.\ (2018), based on four examples fainter than $H_{160}$\,$\sim$\,26 (5\,$\sigma$, assuming a more realistic 1-arcsec source size, rather than a point-source estimate --  comparable to our $K$\,$=$\,25.3 limit);  and $\sim $\,8--20 per cent based on two firm candidates and three tentative sources fainter than $K$\,$=$\,24.8 (5\,$\sigma$) from a parent sample of 25 SMGs in Yamaguchi et al.\ (2019).   We note that all three of these estimates are for sources in the same field:  Franco et al.\ (2018) have sources in common with Simpson et al.\ (2014), although the candidate lists for Yamaguchi et al.\ (2019) and Franco et al.\ (2018) do not overlap. Assuming that the fraction of $K$-faint SMGs does not evolve strongly with submillimetre flux density, which appears to be consistent with the trends in our sample, we would derive a surface density of $K$\,$>$\,25.3 mag and $S_{\rm 870}$\,$\geq$\,1\,mJy SMGs of 450$_{-300}^{+750}$\,deg.$^{-2}$. Wang et al.\ (2019) estimate a surface density of ALMA-detected Extremely Red Objects (EROs, $H_{160}-m_{4.5}$\,$\geq$\,2.5) of $\sim$\,530\,deg.$^{-2}$ at $m_{\rm 4.5}$\,$\leq$\,24, which is in reasonable agreement with our submillimetre-selected sample.   This significant surface density of very red, optically/near-infrared-faint sources is an obvious concern for searches
which rely on similar photometric selection to identify rare examples of high-redshift ``quiescent'' or ``post-starburst'' galaxies (Schreiber et al.\ 2018), as well as for very high redshift Lyman-break galaxies  (Pirzkal et al.\ 2013), especially if these do not allow for very high dust attenuations in the models of foreground contaminating populations.

One of the other basic characteristics of SMGs is the fraction of companions, termed ``multiplicity''  (e.g.\ Ivison et al.\ 2007; Hodge et al.\ 2013; Stach et al.\ 2018), which arises from a combination of physical associations and random projections (exacerbated by the blending in low-resolution single-dish submillimetre surveys, Simpson et al.\ 2020).  For the {\it K-faint} sample of  SMGs we find three examples (0020.1, 0137.1, 0410.1), all fainter than $S_{\rm 870}$\,$\sim$\,2\,mJy, are secondary components  in the fields of other SMGs and another three {\it K-faint} SMGs (0058.0, 0460.0 and 0564.0) are in fields with second fainter SMG.  This rate of multiples is 20 per cent and is identical to the rate in the full sample  (97/496, comprising three triples and 44 pairs). Four of the six {\it K-faint} SMGs have photometric redshift ranges which overlap with the neighbour SMG in their ALMA map (0020.1, 0137.1, 0410.1 and 0546.0).  A simple simulation suggests that such an occurance will happen in $\sim$\,10 per cent of cases by chance, so it is possible that these obscured sources are companions of less-obscured SMGs at the same redshift, illustrating the potential diversity of the population, even at the same redshift  (see Figure~2).  

Now we turn to the properties of the {\it K-faint} SMGs provided by the {\sc magphys} analysis of Dudzevi\v{c}i\={u}t\.{e} et al.\ (2020). In  Figures~2a \& 2b we show the distribution in terms of submillimetre flux density and $K$-band magnitude versus photometric redshift from {\sc magphys} for  the {\it K-faint} SMGs compared to those  with $K$\,$\leq$\,25.3.  These two figures show that the {\it K-faint} subset lie at higher redshifts, but that not all high-redshift SMGs are $K$-faint:  the proportion of {\it K-faint} SMGs at $z$\,$\geq$\,3 is 28\,$\pm$\,4 per cent of the total population, so somewhat higher than the fraction for the total population. This is suggesting that another physical property has to be contributing to the faintness of these sources in the $K$-band, not just their redshifts.  The vectors plotted in Figure~2b indicate the effect of increasing the stellar mass or dust attenuation in the sources, suggesting that the $K$-faint SMGs could be either the lower-mass or higher-attenuation tail of the higher-than-average-redshift SMG population.   One weakness of these plots is the significant uncertainties in  the photometric redshifts, particularly for those sources which are faint in the optical/near-infrared wavebands.  For that reason we show in Figure~2c  the  average redshift Probability Density Functions (PDFs) from the {\sc magphys} analysis of the {\it K-faint} SMGs, compared to the $K$\,$<$\,25.3 subset and also the distribution for the equivalent $K$-faint SMGs from Simpson et al.\ (2014), using the {\sc magphys} modelling results from da Cunha et al.\ (2015).    These distributions confirm  that the {\it K-faint} subset have a significantly higher median redshift compared to the $K$-detected sample: $z$\,$=$\,3.44\,$\pm$\,0.06 versus $z$\,$=$\,2.36\,$\pm$\,0.11. These values are comparable  to those reported for similar selections by Dudzevi\v{c}i\={u}t\.{e} et al.\ (2020).  Equally,  the median redshift we derive for our {\it K-faint}  sample is consistent with  the $z$\,$=$\,3.8\,$\pm$\,0.4 derived for the Simpson et al.\ (2014) examples by da Cunha et al.\ (2015).      Our median redshift is significantly below that claimed for this population by Franco et al.\ (2018), $z$\,$>$\,4, and also the median redshift proposed for the ALMA-detected EROs in Wang et al.\ (2019), $z$\,$=$\,4.0\,$\pm$\,0.2,  but is consistent with the broad range suggested by Yamaguichi et al.\ (2019), $z$\,$\sim$\,3--5.  We note that the expected difference in $K$-band brightness is $\Delta K$\,$\sim$\,1.1\,mag.\ for an SMG at $z$\,$=$\,3.4 compared to $z$\,$=$\,2.4, based on the composite SMG SED from Dudzevi\v{c}i\={u}t\.{e} et al.\ (2020). This difference in typical $K$-band magnitude is consistent with the median trend shown in Figure~2b, as well as the  median $K$ magnitudes of our redshift-matched {\it control} sample, $K$\,$=$\,23.9\,$\pm$\,0.1 mag, compared to the full $K$-detected population, with $K$\,$=$\,22.8\,$\pm$\,0.1 mag.  We conclude that at least part of the explanation for the properties of the $K$-faint SMGs arises from their higher redshifts, compared to the $K$-detected SMG population.  However, as indicated by the two vectors in Figure~2b, it is not clear what other factors may be influencing the $K$-brightness of these SMGs.

The other key parameter which has been historically identified as a driver of extreme red optical/near-infrared colours in star-forming galaxies is dust attenuation, usually parameterised by $V$-band attenuation: $A_V$.  We highlight that the rest-frame $V$-band corresponds roughly to observed $K$-band for SMGs at $z\sim$\,3.5, such as for the {\it K-faint} population.  Using the {\sc magphys} analysis presented in Dudzevi\v{c}i\={u}t\.{e} et al.\ (2020), we derive a median $V$-band attenuation for the {\it K-faint} SMGs of $A_V$\,$=$\,5.2\,$\pm$\,0.3, compared to $A_V$\,$=$\,2.9\,$\pm$\,0.1 for the {\it control} sample, indicating that attenuation, as well as typically higher redshifts, are responsible for the faint near-infrared fluxes of the $K$-faint SMGs -- as suggested by Dudzevi\v{c}i\={u}t\.{e} et al.\ (2020).   We assess the variation in $K$-band flux  for the {\it K-faint}  and   $K$-detected SMG samples versus $z$ and $A_V$ using maximal information-based non-parametric exploration ({\sc mine}, Reshef et al.\ 2011).  This analysis shows that both factors have an influence on near-infrared fluxes on the {\it K-faint} SMGs, which are an order of magnitude fainter than the typical $K$-detected SMG in our survey.   Roughly half of this difference is a consequence of the typically higher redshift of the $K$-faint SMGs  (as suggested by Simpson et al.\ 2014, see also da Cunha et al.\ 2015), with the remainder arising from their higher dust attenuation.

%
%
\begin{figure*}
\centerline{
\psfig{file=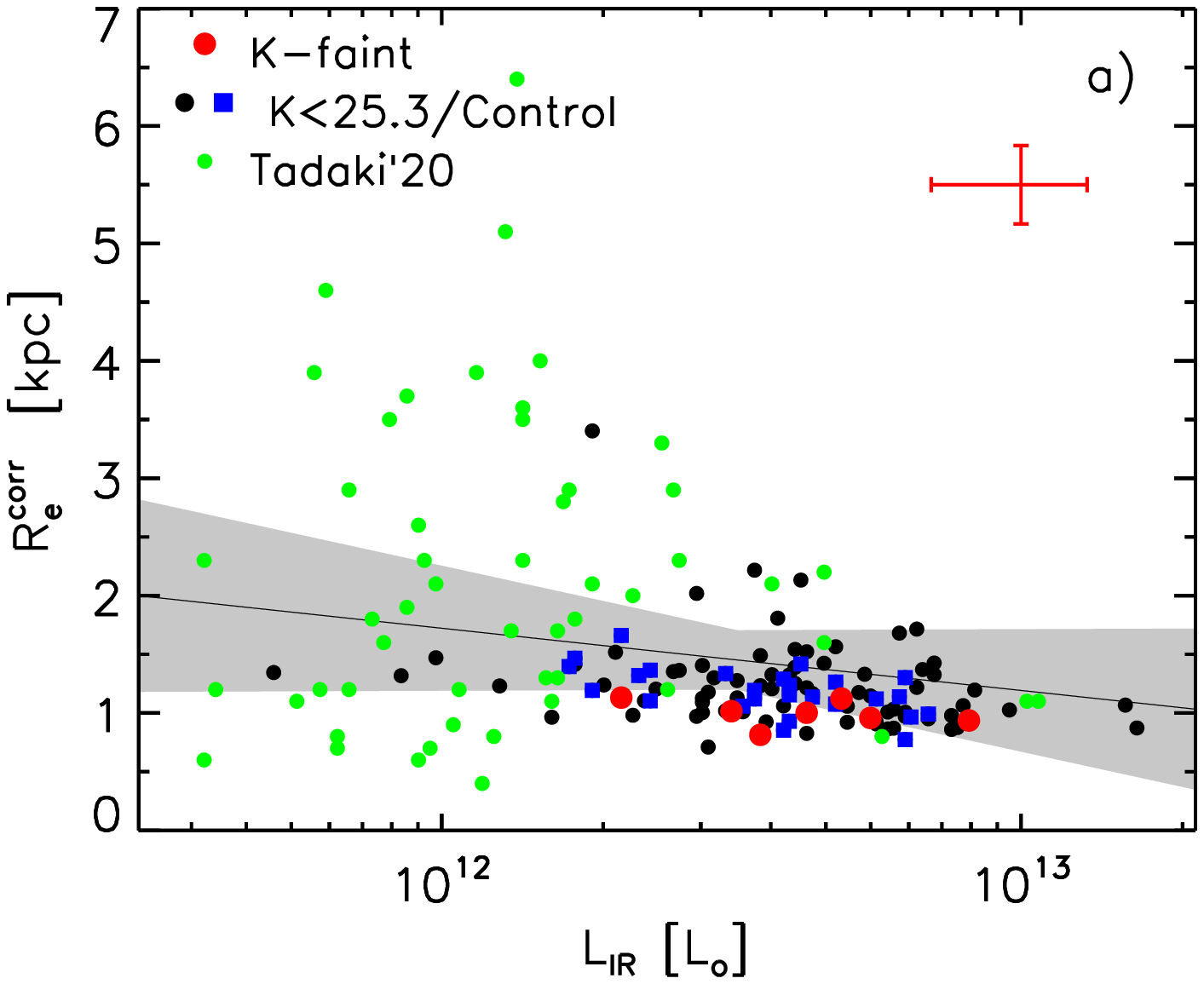,width=3.8in,angle=0}
\hspace*{-0.5in}
\psfig{file=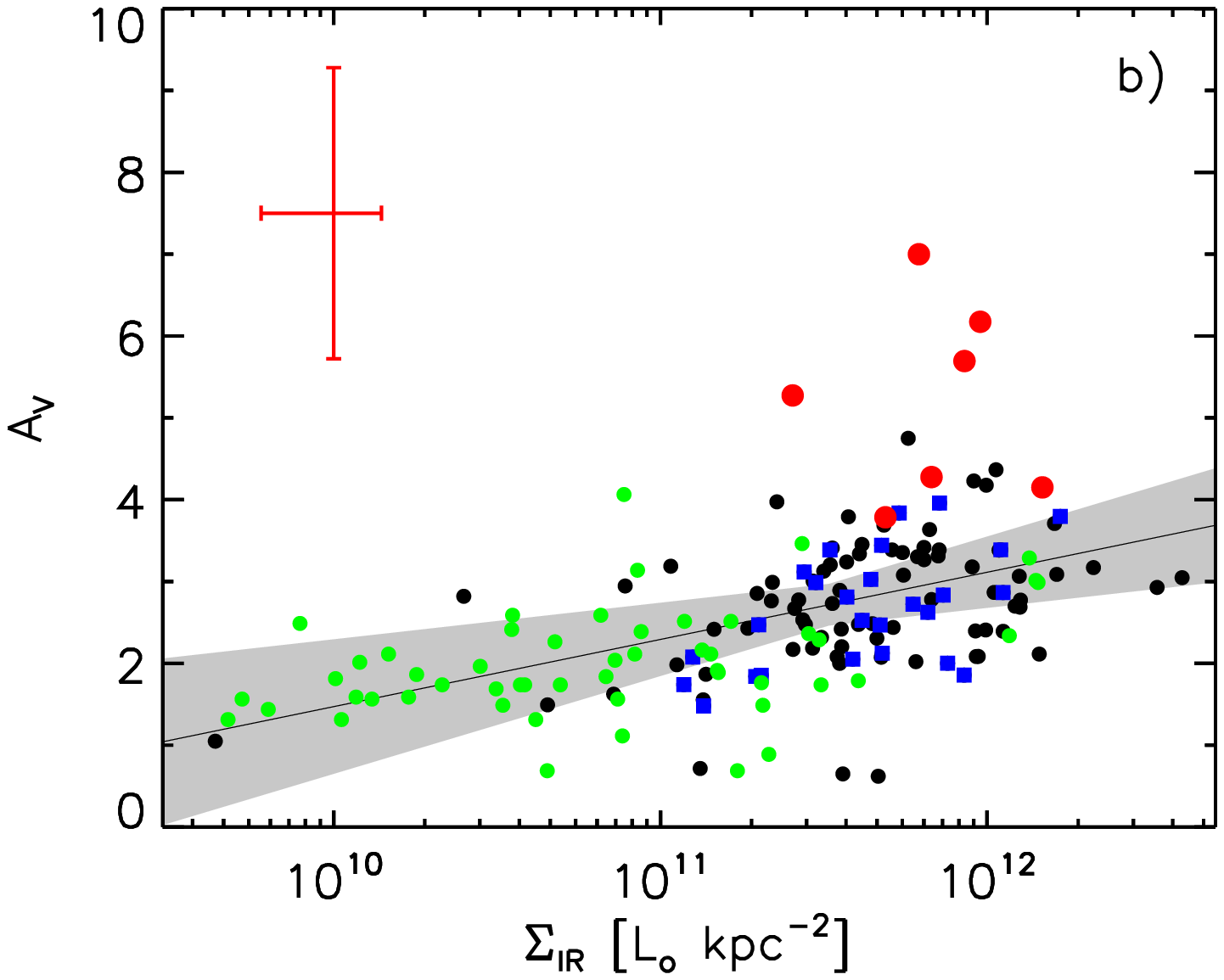,width=3.8in,angle=0}}
\caption{\small 
{\it a)}   Variation of the dust continuum size, $R^{\rm corr}_{\rm e}$, with far-infrared luminosity, $L_{\rm IR}$, for the SMGs from Gullberg et al.\ (2020).  $R^{\rm corr}_{\rm e}$ are the effective radii which have been statistically corrected for the faint, extended component found in stacks of the population as described in \S2.  We identify the {\it K-faint} SMGs and those in our {\it control} sample and we also plot the typically fainter 870-$\mu$m detected $K$-band-selected galaxies from Tadaki et al.\ (2020).  We find a weak trend of increasing size at fainter luminosities (or simply a wider range in sizes), but with considerable scatter,  fitting a linear relation to the combined sample and show this as a solid line and the 1-$\sigma$ uncertainty as the shaded region.    The median uncertainties for the {\it K-faint} SMGs are illustrated by the error bars.
{\it b)}   Variation of the $V$-band attenuation, $A_V$, with far-infrared surface brightness, $\Sigma_{\rm IR}$, for the SMGs from Gullberg et al.\ (2019) and the $K$-band selected sources from Tadaki et al.\ (2020).   We see a correlation between dust attentuation and far-infrared surface brightness (which is a proxy for star-formation rate surface density: $\Sigma_{\rm SFR}$), with the {\it K-faint} SMGs appearing as outliers to the trend, and we fit a linear relation to the combined sample and show this as a solid line and the 1-$\sigma$ uncertainty as the shaded region.    
We use the same symbol as in the left-hand panel and again the median uncertainties for the {\it K-faint} SMGs are illustrated by the error bars.
}
\end{figure*}

While the role of redshift in defining the $K$-faint SMG population is easy to understand, through the distance modulus and $K$ corrections, what is the physical origin of the high dust attenuation seen in these systems, compared to the less extincted, $K$-detected SMGs at similar redshifts seen in Figure~2? We show in  Figures~3a \& 3b the distributions of $A_V$ for the {\it K-faint} SMG sample, compared to both the whole population and the {\it control} sample, as a function of two possible physical drivers: stellar mass ($M_\ast$, following Garn \& Best 2010); and dust mass ($M_{\rm d}$). We caution that the covariance in the parameters in the {\sc magphys} models means that some care needs to be taken when trying to assess the physical significance of any apparent correlations.  So for example the covariant errors between attenuation and inferred stellar age will mean that sources will move diagonally on the $A_V$--$M_\ast$ plane.   Nevertheless, we find no obvious trends with the median stellar masses of the {\it K-faint} and {\it control} samples being  consistent, 10$^{11.10\pm 0.04}$\,M$_\odot$ and 10$^{11.00\pm 0.06}$\,M$_\odot$, as are the median dust masses, 10$^{8.86\pm 0.06}$\,M$_\odot$ and 10$^{8.98\pm 0.04}$\,M$_\odot$.   This suggests that neither dust mass, nor stellar mass, are the drivers of the high dust attenuation of the $K$-faint SMGs.  

As a brief aside, we note that although there have been  few theoretical predictions for the properties of optical/near-infrared faint submillimetre sources, some are provided in the recent work of  Lagos et al.\ (2020) using their {\sc shark} semi-analytic model.  Lagos et al.\ (2020) give  predictions for model SMGs with $S_{870}$\,$>$\,1\,mJy  and $m_{3.6}$\,$>$\,23.5, which roughly match the AS2UDS $K$-faint sample discussed here (see \S2.2).  The predicted properties of these sources in the model  are a median redshift of $z$\,$=$\,3.5$^{+0.9}_{-0.7}$, stellar mass of 10$^{10.0\pm 0.3}$\,M$_\odot$,   dust mass of 10$^{7.5\pm 0.6}$\,M$_\odot$ and attenuation of $A_V$\,$=$\,2.2\,$\pm$\,0.5.  In comparison to the observations results above, we see that the predicted median redshift from {\sc shark} is in good agreement with that for our sample, while the stellar and dust masses appear to be an order-of-magnitude or more too low and the attenuation is $\sim$\,3 magnitudes too low.  Hence the model galaxies tend to be less obscured (indeed, less obscured than the {\it control} sample), but also much less massive and dust rich than the observed population.

Now, returning to the empirical correlations: in Figure~3c we show the relation between $A_V$ and $R^{\rm corr}_{\rm e}$. This shows a moderate correlation with $A_V=$\,(3.2\,$\pm$\,0.2)\,$-$\,(3.0\,$\pm$\,0.9)\,$\times$\,$(R^{\rm corr}_{\rm e}- \left\langle R_{\rm e} \right\rangle)$ with an average $\left\langle R_{\rm e} \right\rangle$\,$=$\,1.13\,kpc for the 33 sources in the combined {\it control} and {\it K-faint} SMG sample with measured sizes. Most striking is that  the {\it K-faint} SMGs fall at the high-$A_V$ and small-$R_{\rm e}$ end of the distributions.  The median dust continuum size of the seven {\it K-faint} and 26 {\it control} SMGs with  sizes are  1.00\,$\pm$\,0.05\,kpc and 1.17\,$\pm$\,0.04\,kpc, with a Kolmogorov-Smirnov test indicating a three per cent chance of the two subsets being drawn from the same parent population.   This suggests that it is the smaller dust continuum sizes which are connected to the high $A_V$, that in turn is  responsible for  the $K$-faint SMGs having the faintest rest-frame optical emission of  SMGs at their redshifts.  The more compact dust emission in the $K$-faint SMGs ought to result in higher characteristic dust temperatures, given the two samples have near identical dust masses (e.g.\ Scoville 2013).  Indeed, this difference in sizes is reflected in a similar difference between the characteristic dust temperatures (assuming the same dust opacity) of the {\it K-faint} and {\it control} samples (for those SMGs with SPIRE detections): $T_{\rm d}$\,$=$\,33.6\,$\pm$\,1.2\,K and  $T_{\rm d}$\,$=$\,30.9\,$\pm$\,1.0\,K, respectively, although this is not statistically significant, amounting to only a 1.7-$\sigma$ difference (see also da Cunha et al.\ 2015). 

To investigate the origin of the trend seen in Figure~3c, we show two plots in Figure~4 which combine dust continuum sizes, dust attenuation and far-infrared luminosities.   Figure~4a illustrates the distribution of dust continuum size with far-infrared luminosity, while in Figure~4b we combine these two parameters to yield the far-infared surface brightness (as a proxy for star-formation rate surface density, $\Sigma_{\rm SFR}$) and plot this against   dust attenuation.  We show both our SMG sample and, to extend the dynamic range of the plots to search for correlations, also the typically fainter dust-continuum-detected sources from Tadaki et al.\ (2020).   We fit log-linear relations to the full distribution of sources in both these plots and derive median trends of
$R_{\rm e}$\,$=$\,(1.45\,$\pm$\,0.14)\,$-$\,(0.53\,$\pm$\,0.31)\,$\times$\,$\log_{10}(L_{\rm IR}/ \left\langle L_{\rm IR} \right\rangle)$,  where $\left\langle L_{\rm IR} \right\rangle$\,$=$\,3.3\,$\times$\,10$^{12}$\,L$_\odot$,  for Figure~4a and  $A_V$\,$=$\,(2.72\,$\pm$\,0.09)\,$+$\,(0.82\,$\pm$\,0.13)\,$\times$\,$\log_{10}(\Sigma_{\rm IR}/\left\langle \Sigma_{\rm IR} \right\rangle)$,  with $\left\langle \Sigma_{\rm IR} \right\rangle$\,$=$\,3.3\,$\times$\,10$^{11}$\,L$_\odot$\,kpc$^{-2}$, in Figure~4b.   The trend of $R_{\rm e}$--$L_{\rm IR}$ in Figure~4a is weak and only marginally significant,  indeed the behaviour might be better described as an increase
in the range of  $R_{\rm e}$ of the population at lower luminosities, although we caution that the statistical correction applied to the AS2UDS sample will not capture the variety of sizes in that sample.  In contrast, the trend we see between 
$A_V$--$\Sigma_{\rm IR}$ is significant ($>$\,6-$\sigma$), indicating that the {\sc magphys}-derived dust attenuation does increase with far-infrared surface brightness (and less strongly with far-infrared luminosity).   We see that the {\it K-faint} SMGs have above-average $\Sigma_{\rm IR}$ for the AS2UDS sample, which  given the measured correlation, would lead to higher attenuation.  However, we also find that the {\it K-faint} SMGs  are indistinguishable in $\Sigma_{\rm IR}$ (or $\Sigma_{\rm SFR}$) from the {\it control} sample, and they  lie off the trend in terms of  $A_V$, indicating that it is not solely their high-surface brightnesses which are responsible for their high attenuation.

We have also studied the trends of dust attenuation with near-infrared size or the ratio of near-infrared/submillimetre size in the $K$-bright AS2UDS SMGs (and the sources from Tadaki et al.\ 2020).  However, there are no clear trends which would explain the properties of the high-$A_V$ inferred for the {\it K-faint} SMGs, which obviously lack detectable near-infrared counterparts and hence sizes.  Thus we cannot conclusively explain the physical origin of the very high dust attenuation seen in this subset of the SMG population.   Nevertheless, there are hints that this behaviour may result from either the relative scale of the obscured and less-obscured components in these systems, or other aspects of their geometry.  Hodge et al.\ (2016) and Gullberg et al.\ (2019) have used high-resolution ALMA maps to demonstrate that the dust continuum emission of SMGs is generally well-described by a Sersic $n$\,$=$\,1 exponential, but the bulk of this emission does not arise from a smooth disk -- instead they trace compact bar-like structures (see Hodge et al.\ 2019), which likely represent dense gas structures driven by external gravitational torques or secularly-generated internal structures in these gas-rich galaxies (the faint spatially-extended exponential halo found by Gullberg et al.\ in their stacks may represent dust emission from the stellar/gas disk).  Thus the smaller sizes we infer for the $K$-faint SMGs  indicates that they host concentrated dust-obscured activity (perhaps lacking a larger disk component), which is at least in part responsible for the inferred high attenuations.  However, the absence of detectable rest-frame optical emission in these systems also suggests that they lack a spatially extended and less-obscured stellar component  (as seen in the optically brighter and lower redshift subset of SMGs, e.g.\ Chen et al.\ 2015).  The absence of an extended, less-obscured stellar component may be an increasingly frequent feature of SMGs at higher redshifts and earlier times where the physical extent of the stellar component in galaxies are expected to be smaller, as there is less opportunity for them to have grown large stellar disks or halos (McAlpine et al.\ 2019).    This combination of vigorous, compact and obscured star formation and an underdeveloped or absent extended unobscured stellar component, suggests that the descendant system will be similarly compact, which has been previously been suggested would naturally link these descendants to the compact ``quiescent'' and ``post-starburst'' galaxies which  exist at $z$\,$\sim$\,1--2, and potentially higher redshifts (e.g., Simpson et al.\ 2014; Toft et al.\ 2014; Hodge et al.\ 2016; Almaini et al.\ 2017).

\section{Conclusions}

We use the large sample of  ALMA-identified SMGs in the S2CLS UDS field from Stach et al.\ (2019), which are covered by ultra-deep $K$-band imaging from UKIDSS UDS ($K$\,$=$\,25.3, 5\,$\sigma$), to construct a sample of 80 $K$\,$>$\,25.3 SMGs with 870-$\mu$m detections of $\geq$\,4.8\,$\sigma$ significance (equivalent to a false positive rate of 1 source).  We select a {\it K-faint} subset of 30 SMGs that are free of potential contamination of their multi-band photometry by nearby galaxies and use these to infer the properties of the larger full $K$\,$>$\,25.3 sample.   In addition, we construct a {\it control} sample of 100 SMGs from the $K$\,$\leq$\,25.3 population, matched in redshift and $L_{\rm IR}$, to allow comparisons free of selection or evolutionary  trends.

Based on our $K$-faint sample, we estimate that 15\,$\pm$\,2 per cent of SMGs brighter than $S_{\rm 870}$\,$\geq$\,3.6\,mJy are fainter than $K$\,$=$\,25.3, rising to $\sim$\,25--30 per cent of those at $z$\,$\gs$\,3.   We see no significant variation in this fraction with 870-$\mu$m flux density across $S_{\rm 870}$\,$=$\,3--10\,mJy.  Hence, extrapolating to the lowest flux densities probed by AS2UDS, we  estimate a surface density of $K$\,$>$\,25.3 and $S_{\rm 870}$\,$\geq$\,1\,mJy SMGs of 450$_{-300}^{+750}$\,deg$^{-2}$.

Using the {\sc magphys} analysis of the AS2UDS SMGs presented in Dudzevi\v{c}i\={u}t\.{e} et al.\ (2020), we investigate the causes of the faintness of these SMGs in the near-infrared and show that the redshift distribution of the {\it K-faint} SMG subset has a significantly higher median  than the $K$-detected SMG subset: $z$\,$=$\,3.44\,$\pm$\,0.06 versus $z$\,$=$\,2.36\,$\pm$\,0.11.

We also show that the median $V$-band attenuation for the {\it K-faint} SMGs of $A_V=$\,5.2\,$\pm$\,0.3, compared to $A_V=$\,2.9\,$\pm$\,0.1 for our redshift-matched {\it control} sample, suggests that a combination of higher attenuation, as well as typically higher redshifts, is responsible for the faint near-infrared fluxes of the $K$-faint SMGs -- as previously proposed by da Cunha et al.\ (2015), Dudzevi\v{c}i\={u}t\.{e} et al.\ (2020) and others.  

Finally, we seek the cause of the high dust attenuation in this subset of SMGs, compared to more average members of the population at the same redshifts.  We find a trend of higher $A_V$ for SMGs with smaller dust continuum sizes, suggesting that it is the compactness of the dust-obscured activity in these systems that is driving the high attenuations we infer. We investigate the relation of dust-continuum sizes with far-infrared luminosity, finding a weak trend for more luminous sources to be more compact, and following on from this the relation between dust attenuation and far-infrared surface brightness, which shows a significant positive correlation.  However, while the higher $\Sigma_{\rm IR}$, and $\Sigma_{\rm SFR}$, values of the {\it K-faint} SMGs indicate higher $A_V$ values, we find that the {\it K-faint} SMGs still lie off this trend.  We suggest
that the behaviour of the $K$\,$>$\,25.3 SMGs reflects details of the relative scales and mixture of dust obscuration and stellar populations in these systems, in particular their properties suggest that any less obscured, spatially-extended stellar component is absent in these galaxies. Testing this will require the deeper, high-resolution mid-infrared imaging which {\it James Webb Space Telescope} ({\it JWST}) can deliver.

We conclude that the $K$-faint SMGs represent the higher-redshift, higher-star-formation-rate-density and higher-dust-attenuation tail of the distribution of the wider SMG population, rather than any distinct subclass or population.    The existance of such systems cautions against using the presence or absence of a near-infrared  counterpart  as a test of the reality of a submillimetre source (as well as highlighting the potential incompleteness of any prior catalogues used to deblended far-infrared observations, if these lack complete interferometric submillimetre identifications), and also highlights the potential for incompleteness in purportedly ``mass''-selected surveys of galaxies at high redshifts and the possibility of contamination in ``drop-out'' selected samples of very high redshift galaxies where the analysis fails to consider the possibility of sources having very high dust attenuation, $A_V$\,$\gg$\,3.  Obviously the upcoming launch of the {\it JWST} holds significant promise for improving our understanding of the nature and properties of these examples of dust-obscured, star-forming galaxies at high redshifts -- including measuring the size of any less-obscured stellar components in the $K$-faint SMG population.

\section*{Acknowledgements}

We thank Cheng Cheng for comments on the paper.
All of the Durham co-authors acknowledge STFC through grant number ST/T000244/1.  
UD acknowledges the support of STFC studentship (ST/R504725/1). 
JEB acknowledges the support of STFC studentship (ST/S50536/1).
CCC acknowledges support from the Ministry of Science and Technology of Taiwan (MOST 109-2112-M-001-016-MY3).
JEG acknowledges support from the Royal Society.
JAH acknowledges support of the VIDI research programme with project number 639.042.611, which 
is (partly) financed by the Netherlands Organisation for Scientific Research (NWO).
JLW acknowledges support from an STFC Ernest Rutherford Fellowship (ST/P004784/1 and ST/P004784/2).
The authors thank John Helly and Lydia Heck for help with HPC and 
we extend our gratitude to the staff at UKIRT for their tireless efforts in ensuring the success of the UKIDSS UDS project.  This research has made use of NASA's Astrophysics Data System (ADS) and the NASA Extragalactic Database (NED). This paper makes use of the following ALMA data: ADS/JAO.ALMA\#2012.1.00090.S, \#2015.1.01528.S, and \#2016.1.00434.S. ALMA is a partnership of ESO (representing its member states), NSF (USA) and NINS (Japan), together with NRC (Canada), MOST and ASIAA (Taiwan), and KASI (Republic of Korea), in cooperation with the Republic of Chile. The Joint ALMA Observatory is operated by ESO, AUI/NRAO and NAOJ.

\section*{Data availability}

The data used in this paper can be obtained from the   UKIRT, {\it Spitzer} and ALMA  data archives, as well as the UKIDSS UDS website: \url{https://www.nottingham.ac.uk/astronomy/UDS}.

\label{lastpage}

\end{document}